\documentclass[referee,a4paper,12pt,traditabstract]{swsc} 

\usepackage{graphicx}
\usepackage{txfonts}
\usepackage{subfigure}
\usepackage{epstopdf}
\usepackage{lineno}
\usepackage[authoryear,round]{natbib}
\usepackage[backref]{hyperref}
\usepackage{url}

\hypersetup{colorlinks=true,citecolor=cyan,urlcolor=cyan,linkcolor=blue}

\begin{document}

   \title{Automated Coronal Hole Identification via Multi-Thermal Intensity Segmentation}
   
   \titlerunning{Automated Coronal Hole Segmentation}

   \authorrunning{Garton, Gallagher and Murray}

   \author{Tadhg M. Garton,
          Peter T. Gallagher
          \and
		  Sophie A. Murray
}

   \institute{School of Physics, Trinity College Dublin, Dublin 2, Ireland\\ 
              \email{\href{gartont@tcd.ie}{gartont@tcd.ie}}
              }


 
  \abstract
   {Coronal holes (CH) are regions of open magnetic fields that appear as dark areas in the solar corona due to their low density and temperature compared to the surrounding quiet corona. To date, accurate identification and segmentation of CHs has been a difficult task due to their comparable intensity to local quiet Sun regions.
   Current segmentation methods typically rely on the use of single EUV passband and magnetogram images to extract CH information.
   Here, the Coronal Hole Identification via Multi-thermal Emission Recognition Algorithm (CHIMERA) is described, which analyses multi-thermal images from the Atmospheric Image Assembly (AIA) onboard the Solar Dynamics Observatory (SDO) to segment coronal hole boundaries by their intensity ratio across three passbands (171~\AA, 193~\AA, and 211~\AA).
   The algorithm allows accurate extraction of CH boundaries and many of their properties, such as area, position, latitudinal and longitudinal width, and magnetic polarity of segmented CHs. From these properties, a clear linear relationship was identified between the duration of geomagnetic storms and coronal hole areas. CHIMERA can therefore form the basis of more accurate forecasting of the start and duration of geomagnetic storms.
   }        


   \maketitle

\section{Introduction}
     \label{S-Introduction} 


The Sun holds a high impact for space weather phenomena through the interactions of solar wind streams with Earth's magnetosphere which can result in geomagnetic storms \citep{Tsurutani06}. Solar wind streams can damage satellites through differential and bulk charging, and geomagnetic storms may interfere with electrical power networks via geomagnetically induced currents \citep{Blake16,Boteler01,Huttunen08,Marshall12}. High speed solar wind streams can be traced back to low density, dark, open magnetic field regions of the corona, known as coronal holes (CHs); \citep{Altschuler72,Antonucci04,Cranmer09,Fujika05}. Here, a new detection algorithm is discussed which segments coronal boundaries from images of the solar corona to ultimately improve predictions of solar wind streams and their properties at 1 AU, and hence, their impact on the Earth's magnetosphere.

\begin{figure}[t]
\centering
\hspace{0pt}
\vspace{0pt}
\includegraphics[width=0.9\textwidth]{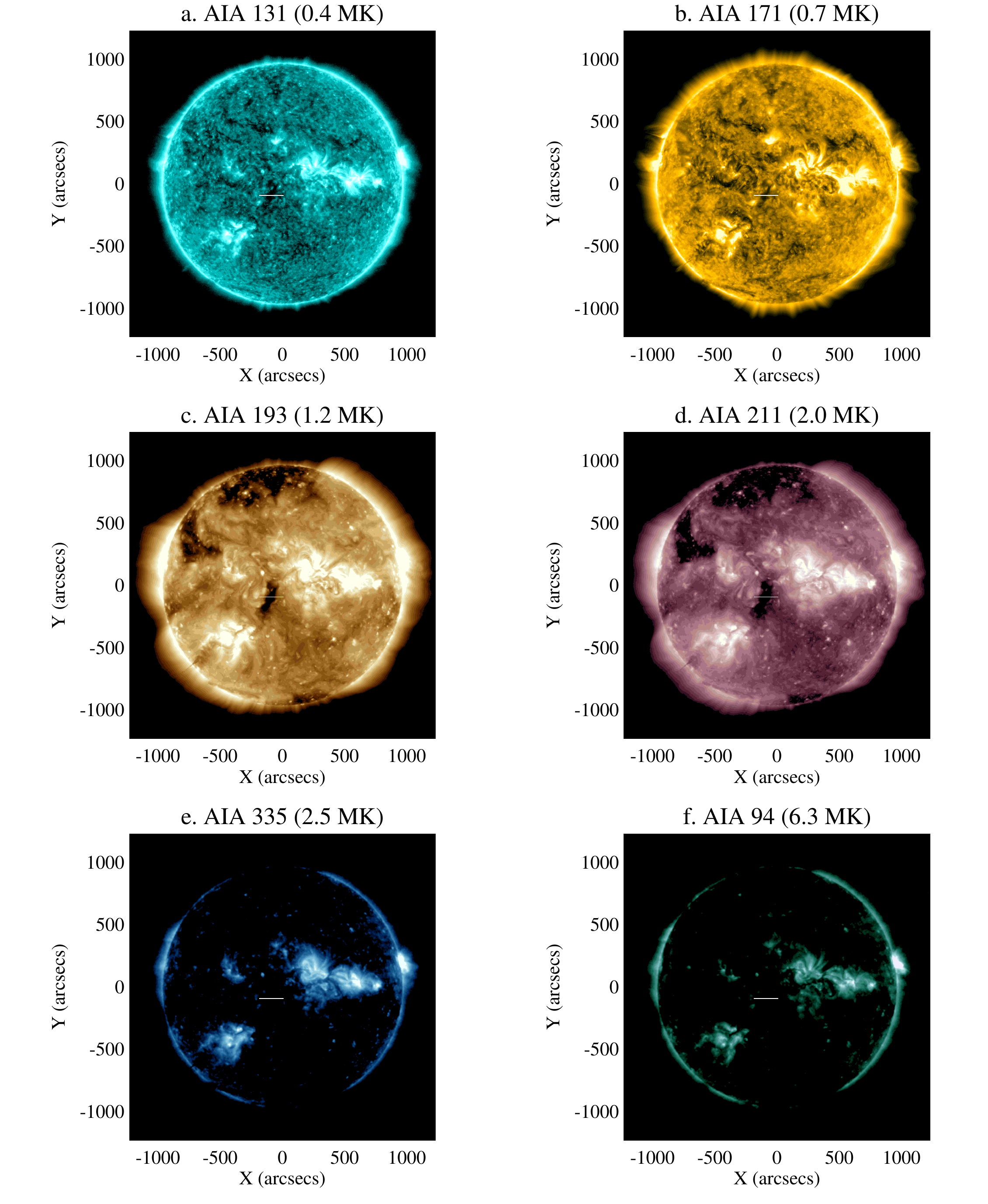}
\vspace{0pt}
\caption[AIA EUV images in six passbands taken of the Sun on 22 September 2016]{AIA EUV images in six passbands taken on 22 September 2016 with a CH located at the central meridian. An intensity cut taken across the CH is shown in Figure~\ref{comp1}.}
\label{sixim}
\vspace{0pt}
\end{figure}

When observing the corona, CHs appear as dark regions on the solar disk and off the solar limb (see Figure~\ref{sixim}), particularly in the 193~\AA\ range due to their low density ($\sim$4$\times$10\textsuperscript{8}~cm\textsuperscript{-3}) and temperatures ($\sim$0.7~MK) compared to the surrounding quiet Sun regions \citep[$\sim$1.6$\times$10\textsuperscript{9}~cm\textsuperscript{-3} and $\sim$1~MK respectively;][]{Phillips95}. CHs, while not visible in any other atmospheric solar layer, are indicators of open magnetic field regions which extend far into the heliosphere \citep{Krieger73}. These extended fields give CHs an ill-defined cut off height and enable CHs to be observed extending far out to the outer limits of the low corona. These high altitude CHs are observed off the solar limb coinciding with near-limb, on-disk CHs, and this phenomena is particularly visible coinciding with polar CHs.

Accurate detection and segmentation of CHs is made difficult by their irregular profile and comparable intensities to the nearby local quiet Sun regions. Detection becomes further difficult  due to off-disk-center CHs often becoming obscured or occulted by brighter, high-density features such as streamers or active region loops. Initially, CH segmentation was done by eye based on two-day averages of He I (10830~\AA) images and magnetograms \citep{Harvey02} taken by the Kitt Peak Vacuum Telescope (KPVT). This method adopted the use of magnetograms to separate CHs from other dark coronal features due to the uni-polar nature of CHs and is still popularly used \citep{Cranmer09}. \cite{Henney05} attempted to automate this detection process and, in turn, created the first automated CH segmentation algorithm. Post-launch of the Solar and Heliospheric Observatory (SOHO), new automated segmentation algorithms were developed that used the clarity of CHs in the 195~\AA\ passband for identification purposes. \cite{Scholl08} implemented the primary multi-passband detection method. Contrast enhanced images in the 171~\AA, 195~\AA, and 304~\AA\ passbands from the Extreme Ultraviolet Telescope (EIT) on-board SOHO were used to differentiate CH `candidate features' from quiet Sun and active regions. These candidate features were then separated into CHs and filaments by comparison with the \cite{Harvey02} He~I coronal hole maps and corresponding magnetograms. The Coronal Hole Automated Recognition and Monitoring (CHARM) algorithm, developed by \cite{Krista09}, uses a histogram-based intensity thresholding technique to differentiate dark coronal regions from quiet Sun regions using only the 195~\AA\ or 193~\AA\ passbands from SOHO or the Solar Dynamics Observatory (SDO) respectively. These dark coronal candidates were separated into CH and non-hole regions through the use of magnetograms. The Spatial Possibilistic Clustering Algorithm (SPoCA; \citeauthor{Verbeeck14}, \citeyear{Verbeeck14}) utilizes fuzzy clustering algorithms and distinctions on area and lifetime of features to differentiate CHs from other features. This algorithm is available for viewing at helioviewer.org. \cite{Reiss15} designed a machine learning algorithm which uses a global intensity thresholding method and the SPoCA algorithm to separate CH features from surrounding regions. Recently, \cite{Boucheron16} developed a detection algorithm which allowed for a varying boundary to search for a maximal change in average intensity between regions inside and outside of this varying contour.

Here, a new method for the detection and segmentation of CHs is discussed. This method is based on multi-thermal intensity segmentation across three EUV wavebands (171~\AA, 193~\AA\ and 211~\AA) from the Atmospheric Imaging Assembly (AIA; \citeauthor{Lemen12}, \citeyear{Lemen12}) on-board SDO \citep{Pesnell12}. This method is used in an automated CH segmentation algorithm, here in named the Coronal Hole Identification via Multi-thermal Emission Recognition Algorithm (CHIMERA). CHIMERA analyses intensities in these three wavebands to estimate the temperature and density of individual pixels and then segments pixels with similar properties to a CH. Next, magnetogram analysis removes smaller non-CH regions that can fall within these constraints by excluding features that don't exhibit the unipolar nature of CHs. In this paper, the stability and accuracy of CHIMERA is demonstrated, and the outputs available for comparison with solar wind measurements at 1 AU are discussed. To categorize a large database of CHs, CHIMERA is automated and runs daily on \url{solarmonitor.org}, providing one CH map per day. 

\section{Observations}
\label{S-Observations}

CHIMERA was developed for use with image sets from AIA and HMI onboard SDO, namely the wavebands: 171~\AA, 193~\AA, and 211~\AA\ and line of sight magnetograms  $\tt{hmi.M\_720s}$. These EUV passbands were centered on the spectral emission lines of FeIX (171~\AA), FeXII (193~\AA), and Fe XIV (211~\AA) and cover a peak temperature range of 7$\times$10\textsuperscript{6} - 2$\times$10\textsuperscript{7} K. Each passband takes 4096$\times$4096 pixel images at a cadence of 12 seconds for AIA and 12 minutes for low noise HMI enabling accurate, time precise feature segmentation. Due to the unique characteristic temperature responses and emission across the AIA wavebands of each coronal feature, it is possible to segment an individual feature by the ratios and magnitudes of emission in each waveband relative to each other. Here, the extraction of CH boundaries from the background coronal features is described by segmenting pixels that exhibit unipolarity, low temperature and low relative density compared to surrounding quiet Sun pixels.

\section{Data Analysis}
\label{S-Data}

\begin{figure}[t]
\centering
\hspace{0pt}
\vspace{0pt}
\includegraphics[width=0.6\textwidth]{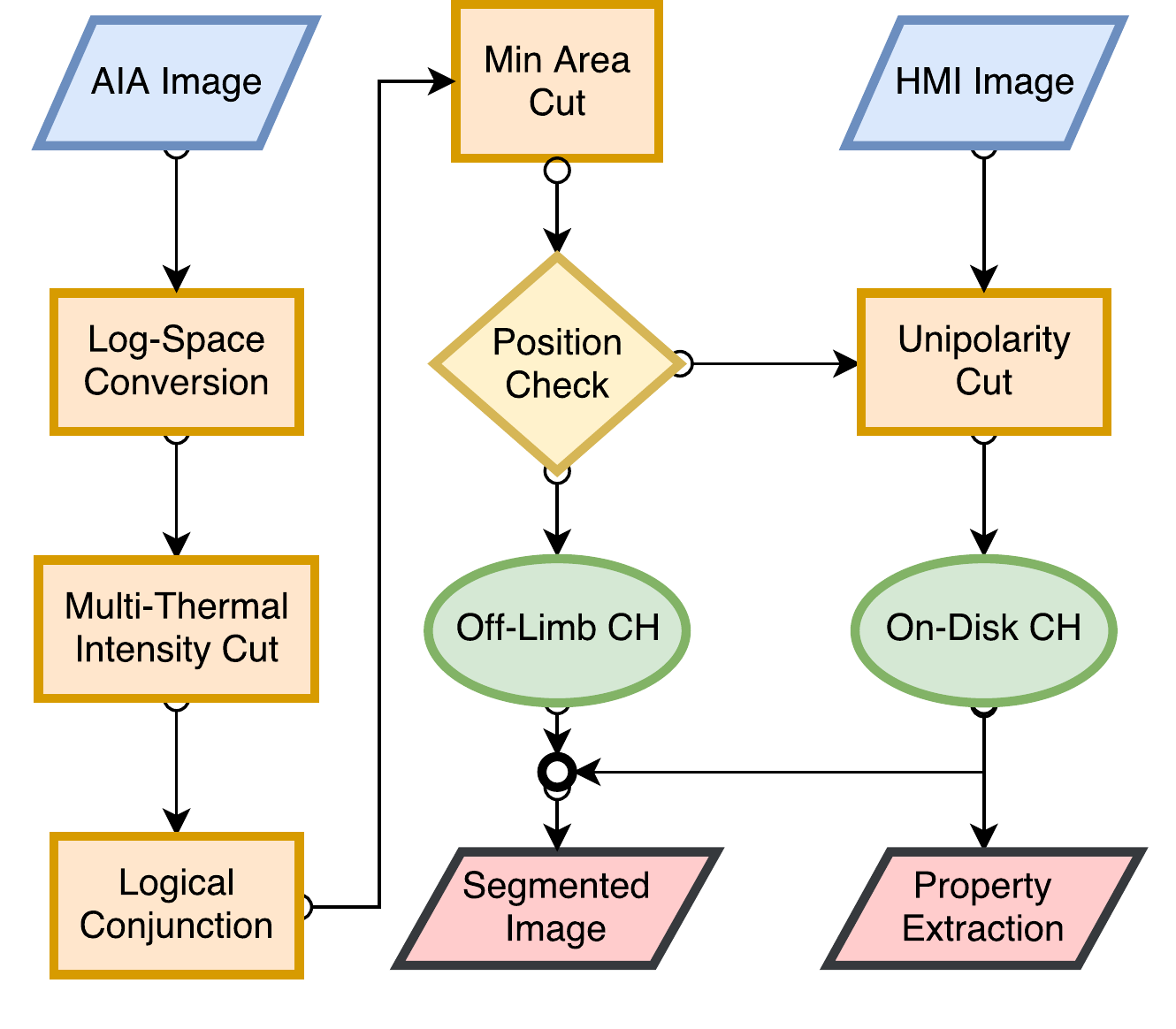}
\vspace{-10pt}
\caption[Flowchart describing the operation for CHIMERA]{A flowchart describing the operation of CHIMERA.}
\label{CHIMflow}
\vspace{0pt}
\end{figure}

\begin{figure}[t]
\centering
\hspace{0pt}
\vspace{0pt}
\includegraphics[width=0.8\textwidth]{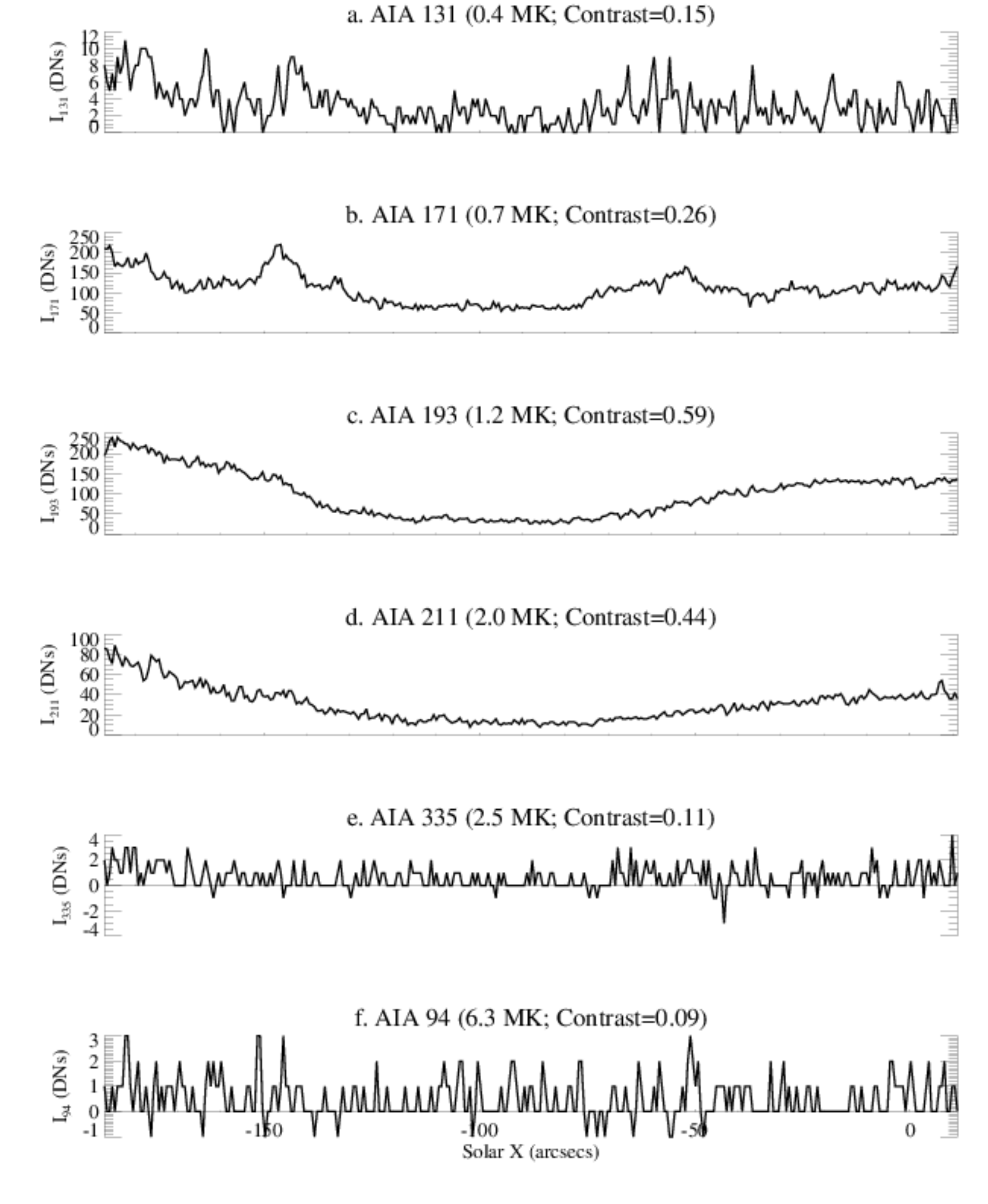}
\vspace{-10pt}
\caption[Intensity cuts across the CH region at central meridian on 22 September 2016]{Intensity cuts across the CH region at central meridian from Figure~\ref{sixim} on 22 September 2016, in units of digital numbers (DNs), ordered by temperature from top to bottom. An approximation of intensity contrast between the CH region and the surrounding quiet Sun, calculated using the Michelson contrast equation (Equation~\ref{contrast}), is displayed in each graph. Difference in intensity between CHs and the surrounding corona is visually noticeable in the 171~\AA\ (b), 193~\AA\ (c), and 211~\AA\ (d) passbands. CHs have a typical temperature of 0.7~MK, which means they should produce peak emission in the 171~\AA\ passband.}
\label{comp1}
\vspace{0pt}
\end{figure}

In this Section, the operation of CHIMERA and the methods used to extract CH boundaries from the three AIA wavelengths (171~\AA, 193~\AA\ and 211~\AA) and one simultaneous snapshot from the Helioseismic and Magnetic Imager (HMI; \citeauthor{Schou12} \citeyear{Schou12}; \citeauthor{Scherrer12} \citeyear{Scherrer12}) is discussed. Figure~\ref{CHIMflow} shows a flowchart describing the segmentation and verification steps of CHIMERA, and is explained further in the following sections:  Multi-thermal intensity segmentation (Section \ref{S-sec2.1}), verification processes (Section \ref{S-maldetect}), and off-limb CH detections (Section \ref{S-sec3.3}). CHIMERA\footnote{See github.com/solarmonitor/solarmonitoridl/blob/master/idl/chimera.pro} was constructed from procedures found in the \textit{SolarSoft} library \citep{Freeland98} with the Interactive Data Language (IDL) programming language.

\subsection{Multi-thermal Intensity Segmentation}
\label{S-sec2.1}

\begin{figure}[t]
\centering
\hspace{0pt}
\vspace{0pt}
\includegraphics[width=0.7\textwidth]{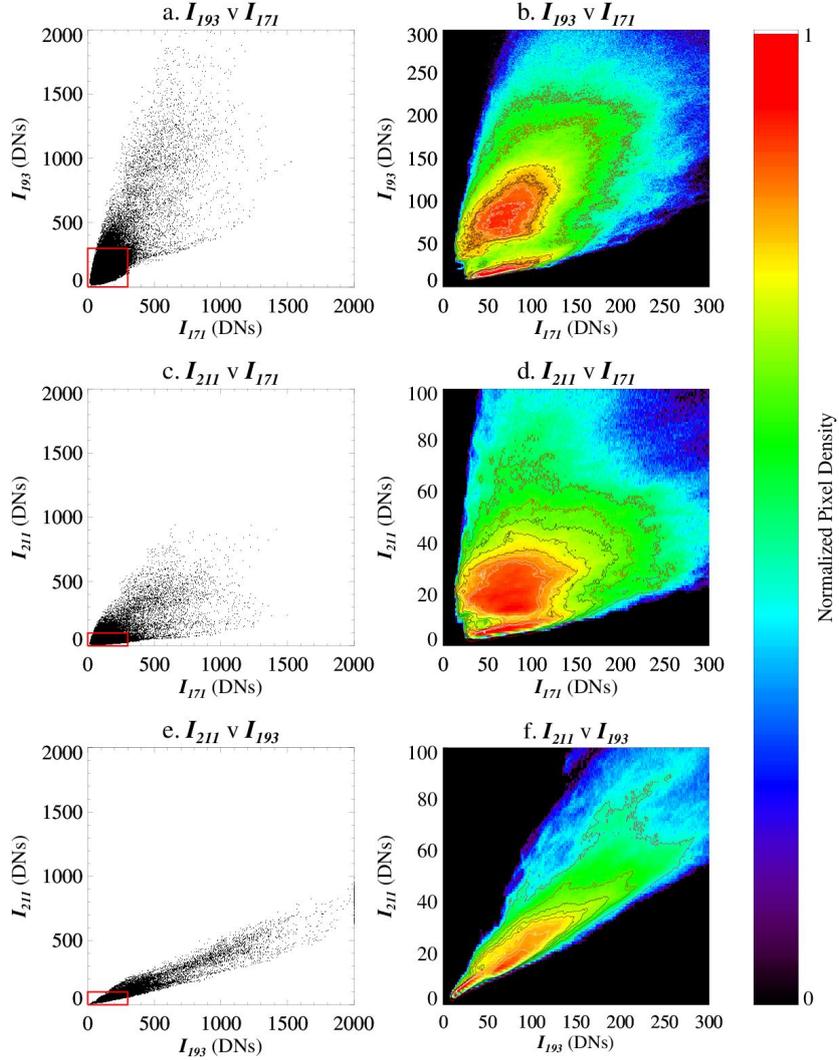}
\vspace{0pt}
\caption[A set of scatterplots and 2D histograms comparing intensities of on disk pixels in 3 wavelengths]{Intensity of on-disk pixels for 31 October 2016 in 193~\AA\ versus 171~\AA\ (a), 211~\AA\ versus 171~\AA\ (c), and 211~\AA\ versus 193~\AA\ (e), with zoom-ins on lower intensities in (b), (d), and (f), respectively. Colour in these zoom-ins denotes the number of solar disk pixels emitting at these wavelengths, with orange/red implying a large number of pixels and purple/blue implying few.}
\label{fig:1}
\vspace{0pt}
\end{figure}

Due to the thermal and density properties of coronal features, each feature has a distinct emission spectra. For example, CHs have a low relative temperature and density, hence they have peak emission in the 171~\AA\ passband and low intensity across other passbands. Figure~\ref{comp1} illustrates an intensity cut across a disk center CH and the contrast, calculated using the Michelson contrast equation (Equation~\ref{contrast}; \citeauthor{Michelson27}, \citeyear{Michelson27}), between this CH to the surrounding corona plasma, which is only visually noticeable in the 171~\AA, 193~\AA\ and 211~\AA\ passbands. 

\begin{equation}
C=\frac{I_{max}-I_{min}}{I_{max}+I_{min}}
\label{contrast}
\end{equation}

\begin{figure}[!t]
\centering
\hspace{0pt}
\vspace{0pt}
\includegraphics[width=0.7\textwidth,height=0.8\textheight,keepaspectratio]{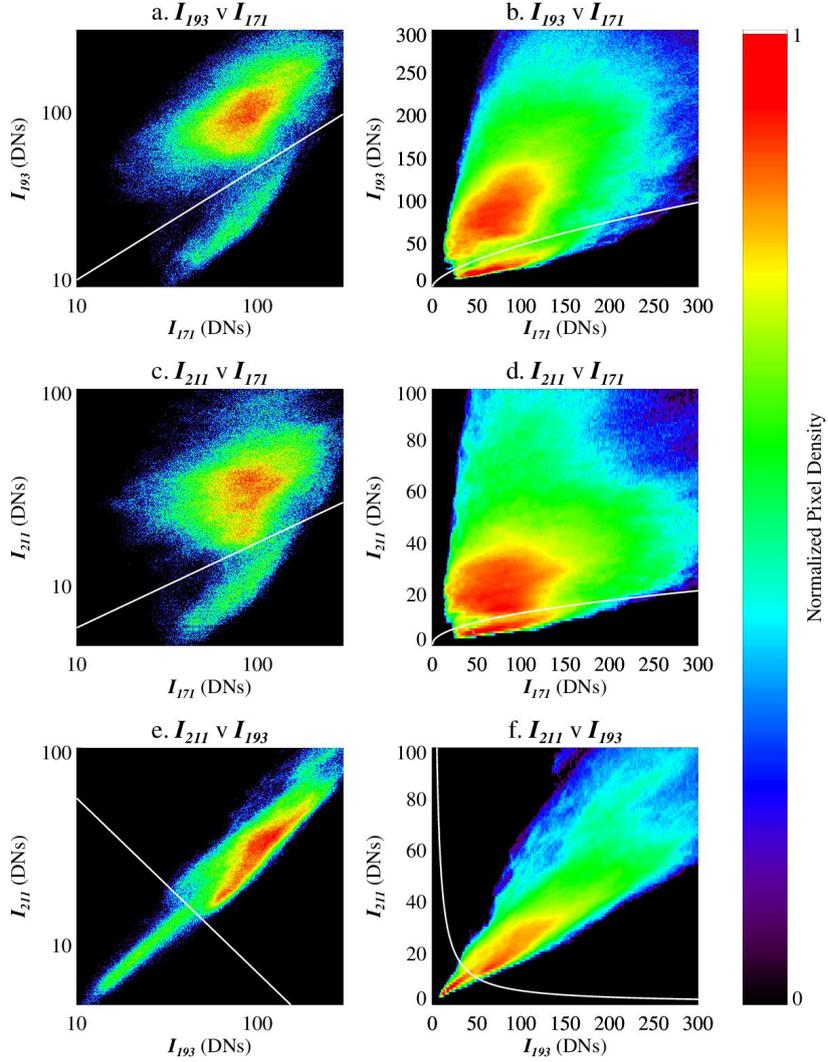}
\vspace{0pt}
\caption[A segmentation curve between the CH and quiet Sun regimes in both log and real space]{Optimum intensity segmentation line fits (white) in log space (a) and intensity space (b) of 193~\AA\ versus 171~\AA, in log space (c) and intensity space (d) of 211~\AA\ versus 171~\AA, and in log space (e) and intensity space (f) of 211~\AA\ versus 193~\AA. Segmentation lines in log-space convert to curves in intensity-space, as described by Equations~\ref{seglin1} and \ref{seglin4}.}
\label{comp}
\vspace{0pt}
\end{figure}

\noindent For this equation, $I_{max}$ is the mean intensity of non-CH pixels near this CH, and $I_{min}$ is the mean intensity of CH pixels within the CH center. By comparing intensities of individual pixels across these three passbands, it is expected for coronal features to cluster by their distinct emission spectra. Figure~\ref{fig:1} (a, c, e) shows three scattergrams comparing the intensities of individual on-disk pixels across these three passbands for a selected date, 31 October 2016. Notably, a high density of pixels, contained within the red boxes in Figure~\ref{fig:1} (a, c, e), exhibit low levels of emission. By taking a close examination of these low intensity pixels, 2D histograms, presented in Figure~\ref{fig:1} (b, d, f), can be created to demonstrate the clustering of points with specific intensities in the analyzed passbands, with red colours implying a higher density of points, and purple/blue implying lower densities. Two regimes appear to exist in each histogram: a large, hotter cluster exists caused by a large number of quiet Sun pixels typically present on the Sun. Smaller clusters also exist, separated from the quiet Sun clusters by valleys, which have a preferred emission in the 171~\AA\ wavelength. This separation of regimes is as expected from the typical temperature and densities of CHs and the ambient corona. Finding an optimum fit to the minimum valleys between the two regimes can be used to separate CH candidates from non-CH regions. This valley is found for $I_{193}$ v $I_{171}$ and $I_{211}$ v $I_{171}$ by finding a number of minima in $I_{y}$ for fixed values of $I_{x}$. Due to the irregular shape of the valley in $I_{211}$ v $I_{193}$, the minima for this instance are instead calculated along lines with slopes parallel to the slope of the dominant intensity-occurrence ridge over the entire $I_{211}$ v $I_{193}$ regime (implemented here as $\sim$$I_{211}$=(0.3)$I_{193}$). Converting these 2D histograms to log-space extends the valleys existing between the clusters, as shown in Figure~\ref{comp} (a,c,e). Fitting a line using a least square regression to the minima of the valley in log-space enables the construction of an optimum segmentation line, as follows in Equations~\ref{seglin1} and \ref{seglin4},

\begin{equation}
\log I_{y}=m \log I_{x}+ \log c
\label{seglin1}
\end{equation}

\begin{equation}
=> I_{y}=c\,I_{x}^{m}
\label{seglin4}
\end{equation}

\noindent where $c=I_{y}(I_{x}=0)$, $m$ represents the slope of a linear equation, $I_{y}$ and $I_{x}$ are variable intensity measurements from a solar image. A segmentation line created by this method allows a variation in both slope and shape in intensity-space, as illustrated in Figure~\ref{comp} (b,d,f). Feature intensities across these wavebands can vary due to instrument degradation, or may vary with the solar cycle. To accommodate these potential variations in intensity, the segmentation line is altered relative to the average intensity of all on-disk pixels for each wavelength. In the case of low intensity in the 193~\AA\ passband the algorithm alters the line to be more stringent with the 193~\AA\ intensities. This is done by multiplying the line (along the $I_{193}$ axis) by: $(<I_{193}>)/(<I_{m}>)$, where $<I_{m}>$ is the measured mean intensity of on-disk pixels for 31 Oct 2016 in the 193~\AA\ passband. This method is similarly performed for the $I_{171}$ and $I_{211}$ measurements. This method gives a simple, rapid, and robust change to the line parameters to continue accurate segmentation of CHs.

\begin{figure}[!t]
\centering
\hspace{-0pt}
\vspace{-0pt}
\includegraphics[width=1.0\textwidth]{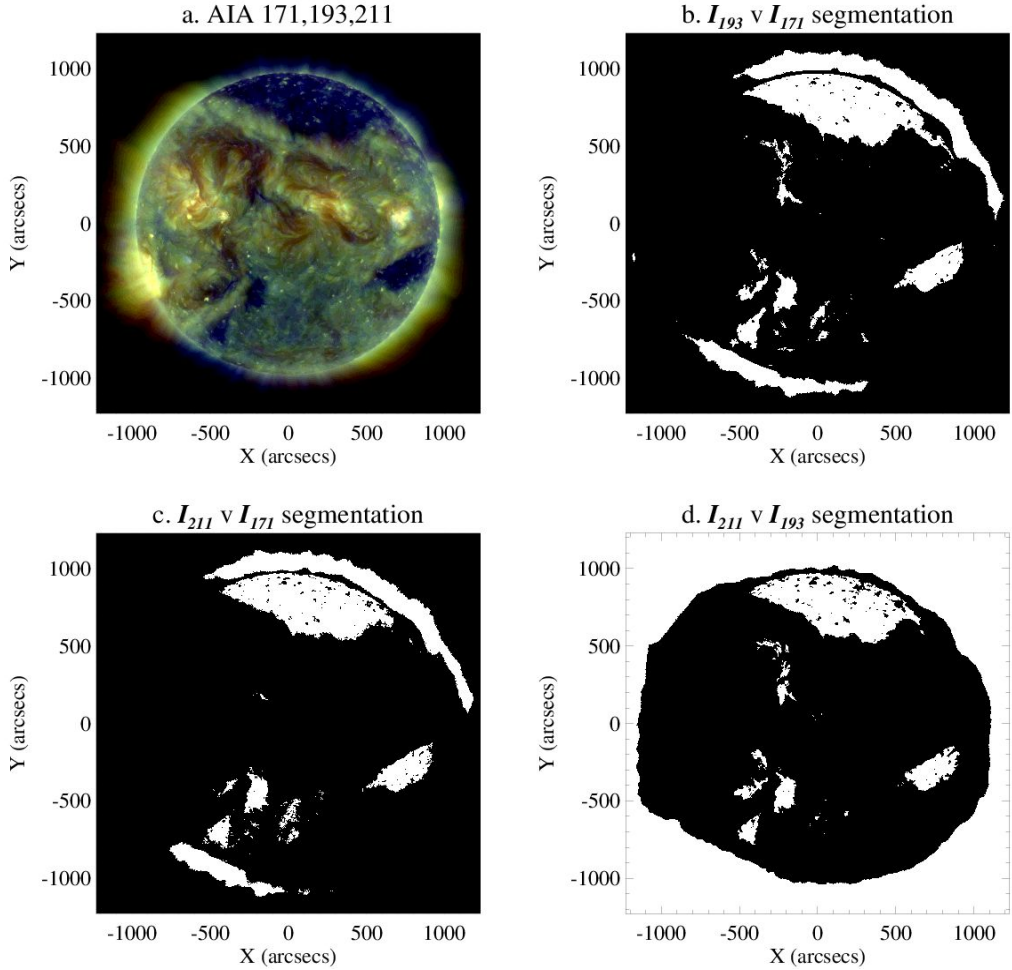}
\hspace{0pt}
\vspace{0pt}
\caption[AIA tri-colour image compared to three CH segmented images]{An example input AIA tri-color image taken on 31 October 2016 in (a), compared to three CH segmentations obtained from the $I_{211}$ v $I_{193}$ (b), $I_{193}$ v $I_{171}$ (c), and $I_{211}$ v $I_{171}$ (d) segmentation curves in Figure~\ref{comp} (b, d, f).}
\label{mask}
\vspace{0pt}
\end{figure}

By taking all regions with intensity ratios below these segmentation curves as CH candidates, three segmented maps are obtained showing all regions that fit these selection criteria. Figure~\ref{mask} (a) displays a tri-color AIA image composed of the 171~\AA\ (blue), 193~\AA\ (green) and 211~\AA\ (red) passbands for 31 October 2016, while the three segmented maps shown in Figure~\ref{mask} (b, c, d) are created by the three segmentation lines from Figure~\ref{comp} (b,d,f). Notably, these maps have similarly segmented regions, however each is inclusive of additional features not present in the other two segmentations, such as a large filament region being detected in both the $I_{211}$ v $I_{193}$ and $I_{193}$ v $I_{171}$ segmentations but being absent in the $I_{211}$ v $I_{171}$ segmentation. Individually, these maps are insufficient to completely segment CH regions from other corona features.

\subsection{Verification Processes}
\label{S-maldetect}

\begin{figure}[!t]
\centering
\hspace{-0pt}
\vspace{-0pt}
\includegraphics[width=1.0\textwidth]{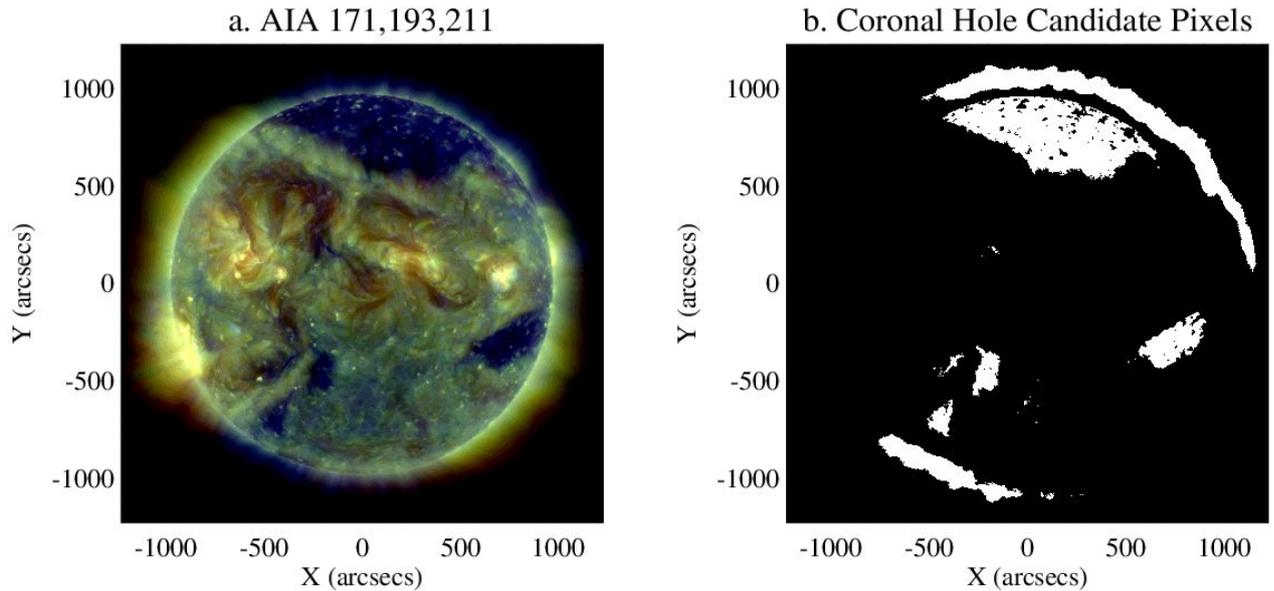}
\hspace{0pt}
\vspace{0pt}
\caption[CH candidates extracted from the logical conjunction of three CH segmentation taken on 31-Oct-2016]{Through the logical conjunction of the three segmentations in Figure~\ref{mask} (b, c, d), CH candidates are obtained from CHIMERA that match the thermal properties exhibited by CH regions on 31 October 2016.}
\label{tmask}
\vspace{0pt}
\end{figure}

As seen in Figure~\ref{mask} (b, c, d), each segmented map is excessively inclusive of CH candidates. However, incorrectly segmented regions are not omni-present in these segmentation maps, while CH regions are. A logical conjunction of the three segmentations removes these extra features while maintaining the detected CH pixels. The output binary map of this process is presented in Figure~\ref{tmask}, having removed the majority of mal-detections while maintaining CH detections.

\begin{figure}[!t]
\centering
\hspace{-0pt}
\vspace{-0pt}

\includegraphics[width=0.9\textwidth]{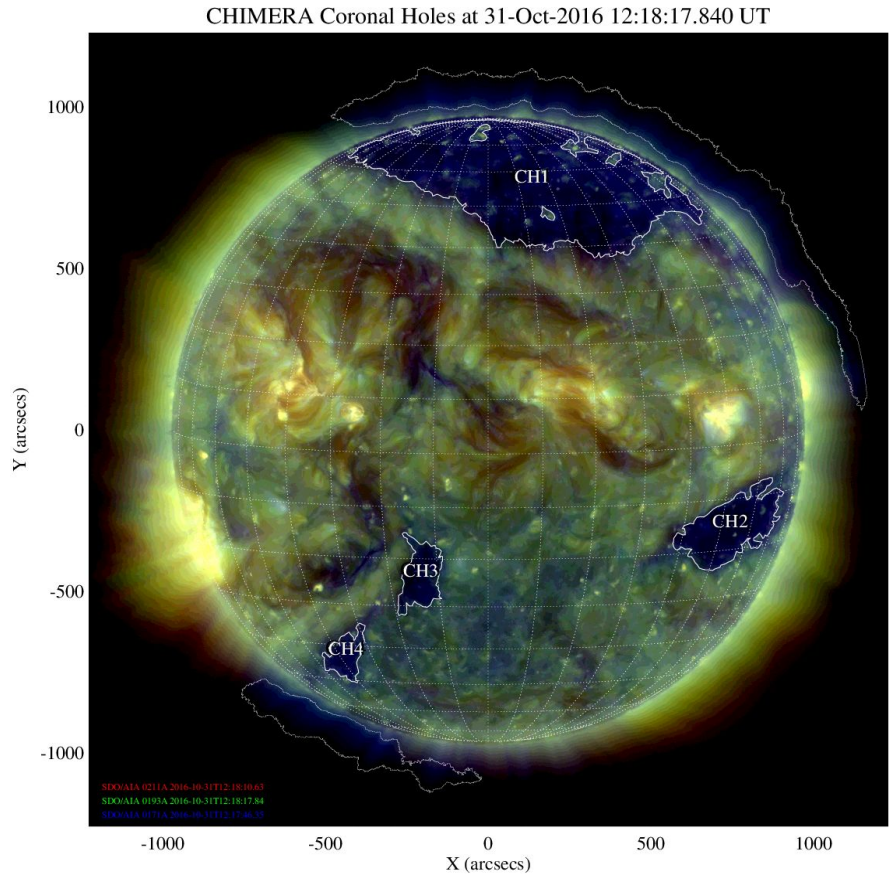}
\hspace{0pt}
\vspace{0pt}
\caption[Output segmented tri-colour image for 31 October 2016]{A segmented image created by CHIMERA for 31 October 2016. CHs regions are extracted from the background quiet Sun and quiet Sun regions trapped within CH boundaries are ignored and removed from detections. Notably off-limb detections, likely to be plumes, are present in this segmented image due to their high correlation with on-disk CH regions and matching thermal properties to CHs.}
\label{seg}
\vspace{-10pt}
\end{figure}

From this newly constructed binary map potential erroneous detections are further excluded by rejecting candidates below 1000 arcsec\textsuperscript{2} ($\sim$30"$\times$30") in size. Small areas could be the result of spurious detections caused by overlapping features or short lived events. This minimum area cut-off follows from the idea of CHs being expansive regions of open magnetic field. This process removes the majority of erroneous detections, however any remaining non-CH regions are removed through the use of HMI magnetograms and the unipolar nature of CHs. This is completed through an approximation of the mean magnetic field of the CH candidate. Any candidates not unipolar in nature are removed from detection, i.e when the mean radial magnetic field strength averaged over the pixels enclosed within the CH candidates boundary is approximated to zero, $<$B$>$\textsubscript{r}$\approx$0~G. The threshold for this cut is calculated relative to the total area of the CH candidate, with the highest stringency being placed on smaller candidates. A small CH candidate is accepted if $<$B$>$\textsubscript{r} is greater than 1~G, and a large CH candidate with an area above 60000 arcsec\textsuperscript{2} is accepted if $<$B$>$\textsubscript{r} is greater than 0.1~G. This varying stringency prevents large candidates being excluded due to a wider range of polarities being present within their boundaries. These thresholds were found from empirical measurements of magnetic properties of the candidates that were present after the previous conjunction and minimum area cut-off. After these final verification steps, segmented CH regions are obtained, as displayed in Figure~\ref{seg}.

\subsection{Off-Limb CH Detections}
\label{S-sec3.3}

In CHIMERA's segmentations some detections exist off the solar limb. These high altitude detections are caused by the open magnetic fields that exist within CH boundaries extending far out into the high corona causing a similar low density and temperature region. These regions, known as plumes, trace out a similar surface shape as that of on-disk CH regions when projected back onto the coronal surface and have matching thermal properties to that of a CH. Furthermore, these plumes are seen to appear off the polar limbs coinciding with polar CHs and off the eastern solar limb pre-appearance of an on disk CH. Thus, it is concluded that CHIMERA is capable of detecting the high altitude component of surface CHs. This enables the use of detected regions off the eastern limb as precursors, which can give insight to the appearance, and potential latitudinal width of a CH, before it rotates onto the near side of the solar disk.

\section{Results}
\label{S-Results}

In this section the accuracy, stability and applications of CHIMERA are discussed, as well as the exact properties calculated for every CH, and potential relations that can be observed between CHs and the effects of the solar wind at 1 AU.

\subsection{Coronal Hole Identification}
\label{S-sec4.1}

\begin{figure}[!t]
\centering
\hspace{-0pt}
\vspace{-0pt}
\includegraphics[width=1.0\textwidth,height=0.87\textheight,keepaspectratio]{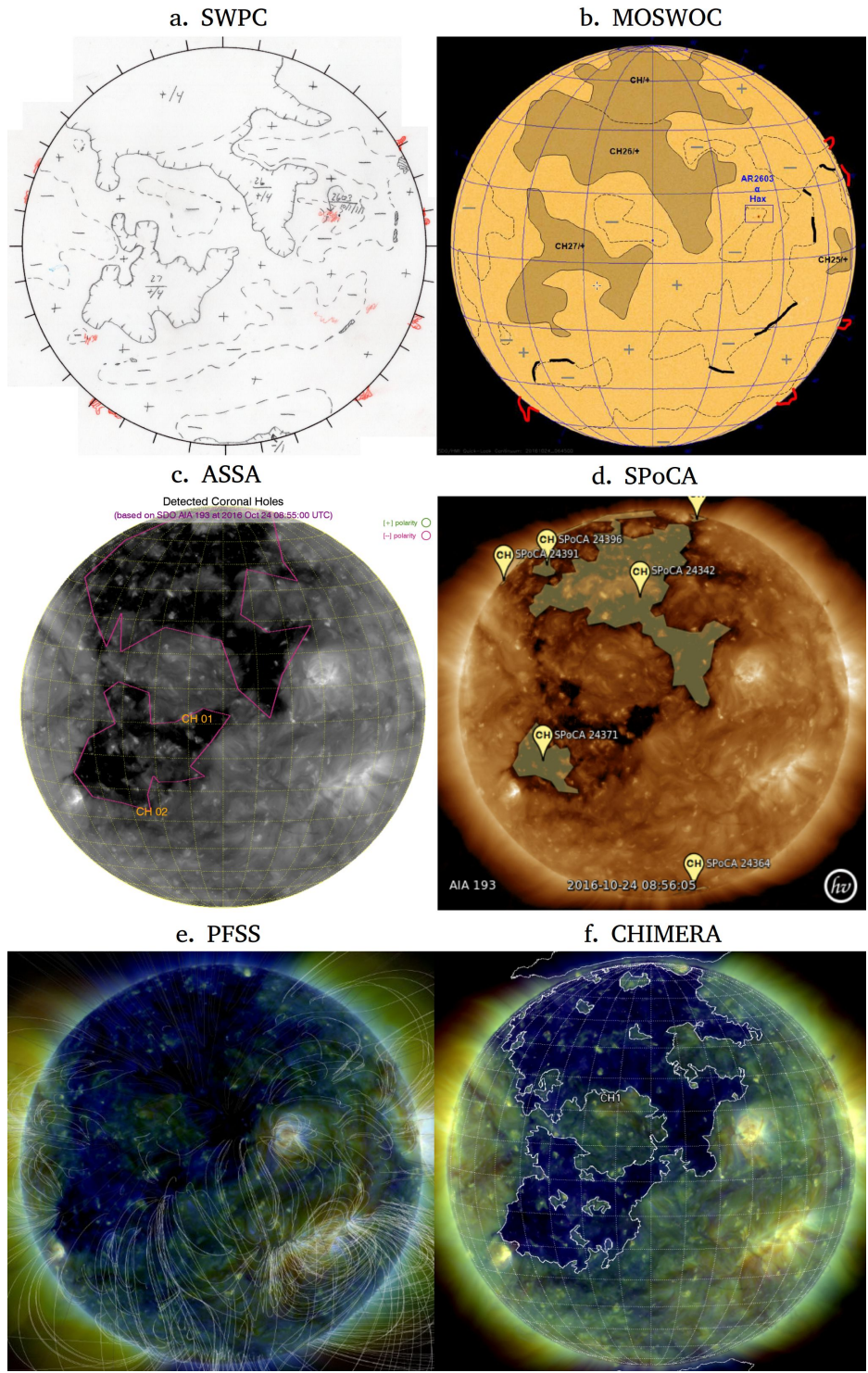}
\hspace{0pt}
\vspace{0pt}
\caption[A comparrison of multiple CH segmentations]{A comparison of CH segmentations for 24 October 2016 output by: (a) NOAA SWPC manually drawn maps, (b) MOSWOC manually drawn maps, (c) ASSA segmented maps, (d) SPoCA segmented maps, (e) PFSS models, and (f) CHIMERA segmented maps.}
\label{segcomp}
\vspace{0pt}
\end{figure}

\begin{figure}[t]
\centering
\hspace{-0pt}
\vspace{-0pt}
\includegraphics[width=1.0\textwidth]{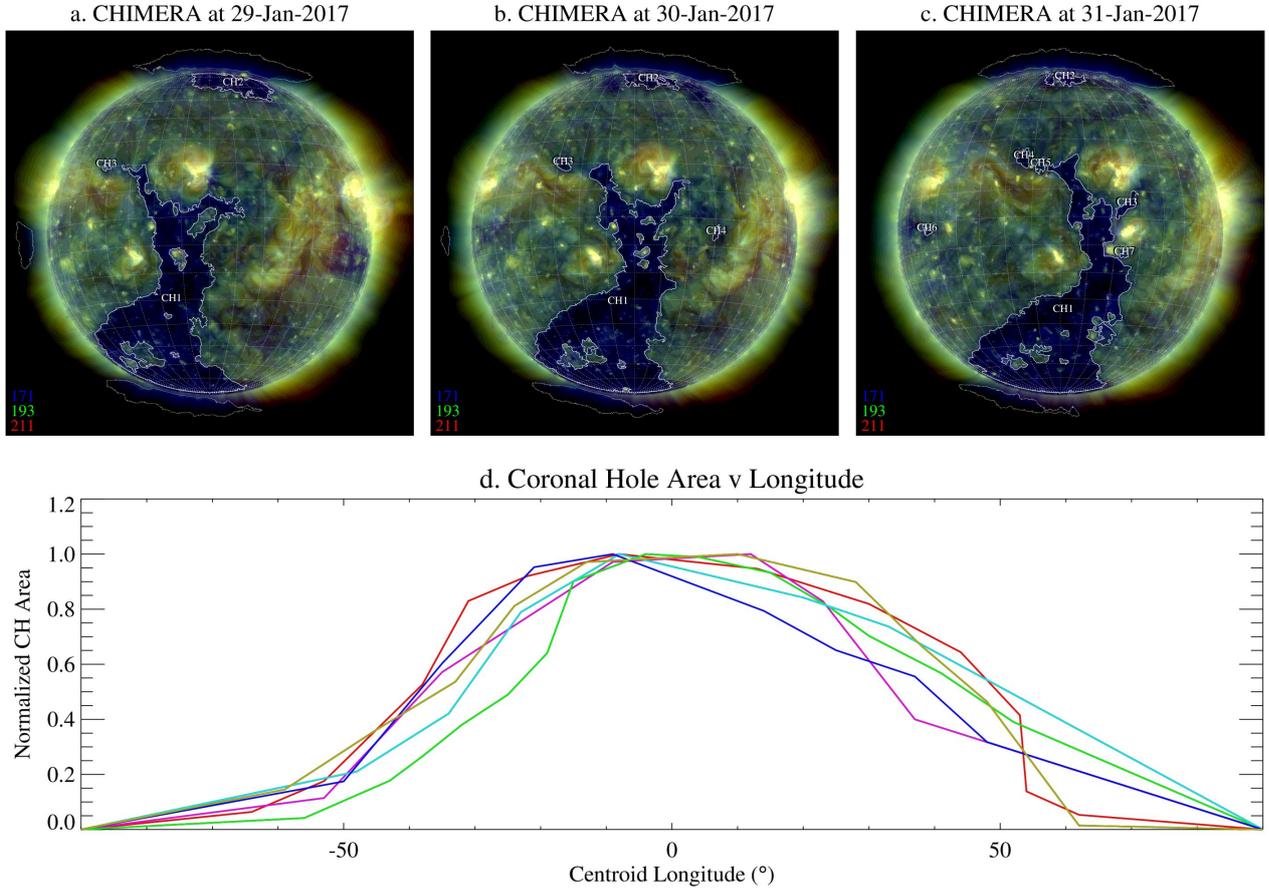}
\hspace{0pt}
\vspace{0pt}
\caption[Three consecutive images demonstrating the stability of CHIMERA's segmentation and tracking abilities]{Three consecutive images taken from 29-31 January 2017 in (a), (b), and (c) demonstrate the short-term stability of CHIMERA's segmentation and tracking abilities. A large geo-effective CH (CH1) is present transversing the central meridian which caused a minor (Kp5o) geomagnetic storm. Mid-term stability is demonstrated in (d), a normalized plot of six randomly selected CH areas compared to their centroid longitude.}
\label{mon}
\vspace{0pt}
\end{figure}

CHIMERA efficiently and effectively extracts CH boundaries in a runtime of $\sim$30 seconds through the use of the 171~\AA, 193~\AA, 211~\AA\ passbands and HMI magnetograms, and the resulting output is shown in Figure~\ref{seg}. The level of accuracy CHIMERA can attain is demonstrated through a comparison between CH boundaries extracted by CHIMERA to other extractions methods for 24 October 2016 in Figure~\ref{segcomp}. This figure displays (a) a manually segmented CH map from the National Oceanic and Atmospheric Administration (NOAA) Space Weather Prediction Center (SWPC)\footnote{www.ngdc.noaa.gov/stp/space-weather/}, (b) a similar manually segemented map from the Met Office Space Weather Operations Center (MOSWOC)\footnote{http://www.metoffice.gov.uk/public/weather/space-weather/}, (c) a segmented map from the Automatic Solar Synoptic Analyzer (ASSA)\footnote{spaceweather.rra.go.kr/models/assa}, (d) a segmented map from the Spatial Possibilistic Clustering Algorithm (SPoCA)\footnote{helioviewer.org}, (e) the result of a Potential Field Source Surface (PFSS)\footnote{www.lmsal.com/~derosa/pfsspack/} extrapolation for comparison of methods with open magnetic field regions, and (f) a segmented map from the CHIMERA\footnote{solarmonitor.org} method of segmentation. As can be seen in this comparison, CHIMERAs detailed segmented CH boundaries follow a similar shape to that of hand-segmented maps, furthermore, CHIMERA visibly better prevents the fracturing of CH detections than that of the SPoCA algorithm. Due to this high level of correlation of CHIMERA's segmentation ability to that of historically accurate segmentation by eye methods, it is concluded that CHIMERA is capable of accurately segmenting CH from the quiet Sun. The stability of CHIMERA's detections are demonstrated in Figure~\ref{mon}, where three solar images segmented by CHIMERA for the consecutive days 29-31 January 2017 (a-c) show the short-term stability of CHIMERA and (d) shows the mid-term stability of CHIMERA by comparing the normalized projected CH area of six randomly selected CH regions between January 2016 to July 2017 compared to their centroid longitude. In this plot small scale variations are due to projection effects of the irregularly shape CH regions. The short-term stability images were selected due to the presence of a large, polar CH (CH1) passing through the central meridian, which caused a minor storm at 1 AU, three days later. This storm registered as a Kp5o storm, according to the World Data Centre for Geomagnetism (WDCG), Kyoto\footnote{wdc.kugi.kyoto-u.ac.jp}. Stability in detections is clearly visible, with each image being a logical one day rotation of the previous with minor boundary morphing expected of a large CH region over a daily timescale.

\subsection{Coronal Hole Property extraction}
\label{S-sec4.2}

From these CH segmentations, CH properties that may hold the highest impact on solar wind properties at 1~AU can be calculated. Table~\ref{t1} illustrates all properties extracted from CH regions by CHIMERA.

\begin{table}[!hb]
\begin{tabular}{ |c|p{12.8cm}| }
	\hline
	Extracted Property & Explanation \\ 
    \hline
    $N_{t}$ & Coronal hole identification number for time $t$. This ID number is assigned in descending order of size.\\
	\hline
    $X/Ycen_{N}$ & Coronal hole centroid coordinates in arcseconds and Stonyhurst heliographic coordinates. \\
    \hline
    $X/Yextent_{N}$ & Most Eastern-Western/Northern-Southern positions in arcseconds. \\
    \hline
    $\Delta\theta$ & Longitudinal angular extent of the coronal hole in Stonyhurst heliographic coordinates and absolute degrees. \\
    \hline
    $A_{tot,N}$ & True coronal hole area. ($\Sigma_{pix}A_{cos,N}$) \\
    \hline
    $A_{\%,N}$ & Percent coverage of the solar disk by the coronal hole area. $\left(\frac{1e(+6){\times}A_{N}}{\pi R_{sun}^{2}}\right)$\\
    \hline
    $<B_{los}>_{N}$ & Mean magnetic polarity along the line of sight. $\left(\frac{\Sigma_{pix}B_{los,N}}{\Sigma_{pix,N}}\right)$\\
    \hline
    $<B^{-/+}_{los}>_{N}$ & Mean negative/positive magnetic polarity for a coronal hole along the line of sight. $\left(\frac{\Sigma_{pix}B^{-/+}_{los}}{\Sigma^{-/+}_{pix,N}}\right)$\\
    \hline
    $B^{min/max}_{los,N}$ & Minimum/maximum magnetic polarity along the line of sight within coronal hole boundaries.\\
    \hline
    $B^{-/+}_{tot,N}$ & Absolute total polarity for all negative/positive pixels within a coronal hole boundary. ($\Sigma_{pix}B^{-/+}_{los,N}$) \\
    \hline
        $<\Phi>_{N}$ & Mean magnetic flux through the surface bounded by the coronal hole boundaries. ($<B_{los}>_{N} A_{tot,N}$)\\
    \hline
    $<\Phi^{-/+}>_{N}$ & Mean negative/positive magnetic flux through the surface bounded by the coronal hole boundaries. ($<B^{-/+}_{los}>_{N} A_{tot,N}$)\\
    \hline
\end{tabular}
\caption{CH properties extracted by CHIMERA.}
\vspace{-10pt}
\label{t1}
\end{table}

\begin{figure}[t]
\centering
\hspace{40pt}
\vspace{-10pt}
\includegraphics[width=1.0\textwidth]{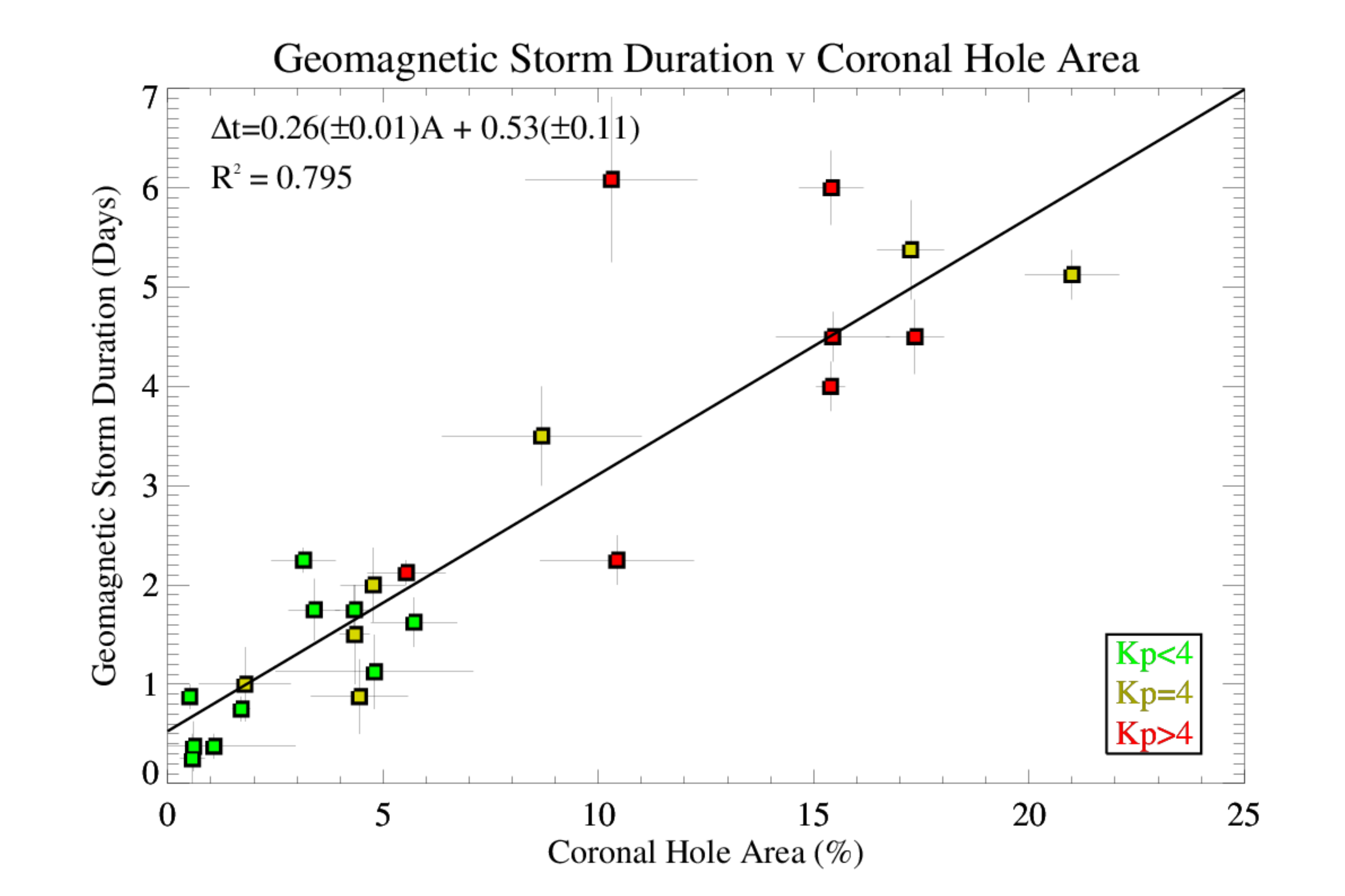}
\vspace{10pt}
\caption[Scatterplot showing the relationship between CH area and geomagnetic storm duration]{A comparison of measurements of CH area, the duration of the geomagnetic storms they produced and their peak Kp index taken from the WDCG. The best fit equation is displayed in the upper left, where $\Delta$t is the geomagnetic storm duration in days, and A is the coronal hole area in percent of the solar disk.}
\label{swcomp}
\vspace{0pt}
\end{figure}

Each property extracted by CHIMERA gives some insight into the current and potential geo-effectivity of a CH. CH area gives possible estimation of the high speed solar wind duration \citep{Krista09,Krieger73} and velocity \citep{Nolte76}, CH extent and positioning can give further insight into duration, as well as arrival time of solar wind streams \citep{Cranmer02}, and magnetic polarity and flux are commonly associated with potential geo-effectivity of solar wind streams. Making these CH properties readily available outputs of CHIMERA allows a statistical analysis of the detected CHs and the solar winds they create, an example of which is displayed in Figure~\ref{swcomp}.

This figure draws a relation between CH area, geomagnetic storm duration and peak storm intensity, taken from the WDCG. Displayed is a best fit line, $\Delta$t = 0.26$\pm$0.01A + 0.53$\pm$0.11, correlating these two properties where $\Delta$t is the geomagnetic storm duration in days and A is the coronal hole area in percent of the solar disk. From this relationship estimations can be made on the geomagnetic storms a given coronal hole will produce, and, as expected, larger coronal holes will typically produce longer geomagnetic storms.

CHIMERA is however not without limitations, like all automated systems. A predominant weakness in CH detection algorithms is the inability to detect CH regions when occulted by brighter, denser regions. Due to pixel intensities being an integration along the line of sight, pixels observing across equal projection of CH and quiet Sun regions will be seen as quiet Sun by CHIMERA. Normally this phenomenon is of little to no importance, but due to CHs being a low density feature they are more easily occulted than other coronal features. Furthermore, with solar wind properties having a high correlation with CH extent and area, inaccurate estimations can cause incorrect estimations of geo-effectivity of CHs.

\section{Conclusions}
\label{S-Conclusions}

A new fast ($\sim$30 second runtime) and robust CH identification technique has been developed which uses HMI magnetograms and the three AIA passbands 171~\AA, 193~\AA, and 211~\AA\ to segment CH boundaries. The algorithm classifies every pixel in each image by the exhibited thermal and density properties across these 3 passbands. This unique classification method can be further extended to classify other coronal features such as filaments, which are similarly dark to CHs but have a peaked emission in the 211~\AA\ passband and are bipolar in nature. Upon completion of segmentation, CHIMERA calculates the centroid, area, latitudinal and longitudinal extent, CH percentage coverage, and many other properties for comparison with solar wind measurements at 1 AU and predictions of upcoming events up to the following 10-14 days.

The primary limitation of this segmentation method is classification of partially or fully occulted CHs. Low density and dim coronal features are highly susceptible to occultation from brighter features, this is particularly prevalent closer to the solar limb where occultation is more common. This effect implies a CH can seem to disappear before it reaches the solar limb, which can cause a CH to become undetectable 1-2 days before expected. These effects can cause errors in boundary segmentation and the calculation of near-limb CHs properties, however, it is currently impossible to accommodate these effects without the use of multiple viewpoints. Unfortunately, as of yet, there are no multiple viewpoints of the Sun in the three used passbands, although, this method of segmentation could be extended to other passbands which exhibit similar contrasts between CHs and other coronal features.

CHIMERA can be used to link properties of on-disk CH regions with the properties of their high speed solar wind streams and the geomagnetic storms they produce. From the CH properties extracted by CHIMERA and measurements of geomagnetic activity it is possible to draw correlations between these two phenomenon. Figure~\ref{swcomp} shows one possible comparison of CH properties, calculated by CHIMERA, to the properties of the geomagnetic storms they produced. These comparisons can be extended further through machine learning algorithms to draw more relations between CH properties and the properties of the solar wind streams they produce.

Further study is required on the morphology of CHs and their interplanetary effects via the high-speed solar wind. Following on from this work, the multi-thermal analysis method of classification can be applied to other coronal features, allowing the segmentation and detection of filament regions or complex active regions and loop structures. Furthermore, outputs of feature properties can be compared statistically with solar wind effects at 1 AU to improve space weather forecasting methods. With the completion of this algorithm, the future aim for this research is to combine this CH detection software with solar wind models to create a real time emulation of solar CHs and the solar wind they produce up to $\sim$1 AU. This combined with statistical analyses of solar wind properties will enable accurate predictions of solar wind arrival and lifetimes of geomagnetic storms affecting the Earth's magnetosphere. Furthermore, this method of segmentation will be updated to segment pixels using 3D intensity histograms and tested on GOES-16 X-ray and EUV images in the future.

\begin{acknowledgements}
T. M. Garton is supported by a Government of Ireland Studentship from the Irish Research Council (IRC). S. A. Murray is supported by the European Union Seventh Framework Program under grant agreement No. 606692 (HELCATS project) and the European Union Horizon 2020 program under grant agreement No. 640216 (FLARECAST project). The editor thanks Véronique Delouille and an anonymous referee for their assistance in evaluating this paper.
\end{acknowledgements}


\bibliography{swsc}

\end{document}